\documentclass[12pt]{article}
\def\beq{\begin{equation}}
\def\eeq{\end{equation}}
\def\ba{\begin{eqnarray}}
\def\ea{\end{eqnarray}}
\def\ga{\mathrel{\raise.3ex\hbox{$>$\kern-.75em\lower1ex\hbox{$\sim$}}}}
\def\la{\mathrel{\raise.3ex\hbox{$<$\kern-.75em\lower1ex\hbox{$\sim$}}}}
\def\Mbh{M_{\rm BH}}
\def\Tbh{T_{\rm BH}}
\def\tbh{\tau_{\rm BH}}
\def\xp{x_{\phi}}
\def\Yp{Y_{\phi}}
\def\mp{m_{\phi}}

\def\tp{\tau_\phi}
\def\GeV{{\rm GeV}}
\def\Mpl{M_{\rm Pl}}
\def\mpl{m_{\rm Pl}}
\renewcommand{\thefootnote}{\fnsymbol{footnote}}
\begin{document}

\mbox{ }\vskip2cm

\begin{center}
%\twocolumn[\hsize\textwidth\columnwidth\hsize\csname @twocolumnfalse\endcsname

%\draft

%\title{
{\bf\Large 
Moduli constraints on primordial black holes}
\vspace{2cm}

%\author{
Martin Lemoine\footnote{{\tt email: Martin.Lemoine@obspm.fr}}
%}

%\address{
{\it
DARC, UMR--8629 CNRS, Observatoire de Paris-Meudon, F--92195
Meudon C\'edex, France}
\end{center}
\vspace{2cm}

%\date{\today}
%\maketitle

\noindent
{\bf Abstract.}  
%\begin{abstract}
The amount of late decaying massive particles (e.g., gravitinos,
moduli) produced in the evaporation of primordial black holes (PBHs)
of mass $\Mbh\la10^9\,$g is calculated. Limits imposed by big-bang
nucleosynthesis on the abundance of these particles are used to
constrain the initial PBH mass fraction $\beta$ (ratio of PBH energy
density to critical energy density at formation), as: $\beta\la
5\times10^{-19} (\xp/6\,10^{-3})^{-1} (\Mbh/10^9\,{\rm g})^{-1/2}
(\overline{\Yp}/10^{-14})$; $\xp$ is the fraction of PBH luminosity
going into gravitinos or moduli, $\overline{\Yp}$ is the upper bound
imposed by nucleosynthesis on the number density to entropy density
ratio of gravitinos or moduli. This notably implies that such PBHs
should never come to dominate the cosmic energy density.

%\end{abstract}

\clearpage
%]
%\noindent{\bf 1. Introduction}
%\vspace{0.3cm}

\renewcommand{\thefootnote}{\fnsymbol{footnote}}
%\narrowtext

{\it 1. Introduction --} The spectrum of locally supersymmetric
theories generically contain fields whose interactions are
gravitational, and whose mass $\mp\sim{\cal O}(100\,\GeV)$. The
Polonyi and gravitino fields of supergravity theories, or the moduli
of string theories, are such examples. This leads to well-known
cosmological difficulties: quite notably, such particles (hereafter
generically noted $\phi$ and termed moduli) decay on a timescale
$\tp\sim \Mpl^2/\mp^3 \sim 10^8\,{\rm s}\,(\mp/100\,\GeV)^{-3}$, i.e.,
after big-bang nucleosynthesis (BBN), and the decay products may
drastically alter the light elements abundances~\cite{W82}. The
success of BBN predictions provides in turn a stringent upper limit on
the number density to entropy density ratio ($\Yp$) of these moduli,
generically $\Yp\la10^{-14}$~\cite{ELN82} (see Sec.~3).

It is argued in this letter that these same constraints can be
translated into stringent constraints on the abundance of primordial
black holes (PBHs) with mass $\Mbh\la10^9\,$g. In effect, moduli are
expected to be part of the Hawking radiation of an evaporating black
hole as soon as the temperature of the black hole exceeds (roughly
speaking) the rest-mass $\mp$; and indeed, the Hawking temperature of
a PBH reads $\Tbh \equiv\mpl^2/\Mbh\simeq 10^4\,\GeV\,(\Mbh/10^9\,{\rm
g})^{-1}$~\cite{H74}.

Primordial black holes are liable to form in the early Universe at
various epochs, e.g., when a density fluctuation re-enters the horizon
with an overdensity of order unity~\cite{ZN67}, or when the speed of
sound vanishes~\cite{J97} (as may occur in phase transitions). As a
consequence, constraints on the abundance of PBHs can be translated
into constraints on the structure of the very early
Universe~\cite{CMcG98}. Until recently, the only existing constraint
on PBHs of mass $\Mbh\la10^9\,$g relied on the assumption that via
evaporation, PBHs leave behind stable Planck mass
relics~\cite{BCL92}. However, recent work from the perspective of
string theories seems to indicate that this is not the
case~\cite{DV99}, in particular that evaporation proceeds
fully. Nevertheless, Green~\cite{G99} has pointed out recently that
such PBHs would also produce supersymmetric particles, and
consequently, cosmological constraints on the lightest supersymmetric
particle (LSP) density could be turned into constraints on the initial
PBH mass fraction $\beta$ (defined as the ratio of PBH energy density
to critical energy density at formation). This constraint relies on
the assumption that the LSP is stable, i.e. $R-$parity is a valid
symmetry; and, as attractive as $R-$parity is, it is not of a vital
necessity altogether. The constraint related to the production of
gravitinos or moduli, to be derived below, is thus complementary to
this $R-$parity constraint, and it also turns out to be more
stringent. Hereafter, units are $\hbar=k_{\rm B}=c=1$, and
$\mpl\equiv\Mpl/(8\pi)^{1/2} \simeq2.4\times10^{18}\,\GeV$ is the
reduced Planck mass.
\bigskip

{\it 2. Moduli production --} Although one is generally interested in
$\Yp$ itself, and not in its momentum dependence, it will prove
necessary in a first approach to keep track of $d\Yp/dk$ (where $k$ is
the momentum) integrated over the black hole lifetime. In effect,
during their evaporation, PBHs produce moduli over a whole spectrum of
momenta, with high Lorentz factors, and the existing constraints on
$\Yp$ depend strongly on the (cosmic) time at which moduli decay
($\tp$ is the decay timescale in the modulus rest frame), hence on
whether they are relativistic or not.

More quantitatively, the mass and temperature of a PBH evolve with
time $t$ during evaporation as: $M(t)=\Mbh\left[1 -
(t-t_i)/\tbh\right]^{1/3}$ and $T(t)=\Tbh\left[1 -
(t-t_i)/\tbh\right]^{-1/3}$~\cite{H74}. Here, $t_i$ denotes the time
of formation, $t_i\ll\tbh$, with $\tbh$ the PBH lifetime: $\tbh\simeq
0.14\,{\rm s} (\Mbh/10^9\,{\rm g})^3$~\footnote{ The lifetime of a
black hole depends on the number of degrees of freedom $g_s$ in each
spin $s$ in the radiation~\cite{McGW90}, {\it i.e.} $\tbh=6.2\,{\rm
s}\, f(\Mbh)^{-1}(\Mbh/10^9\,{\rm g})^3$, with $f(\Mbh)\simeq0.267g_0
+ 0.147g_{1/2} + 0.06g_1 + 0.02g_{3/2} + 0.007g_2$. Here the particle
content of the minimal supersymmetric standard model (MSSM) with
unbroken supersymmetry has been used, $g_0=98$, $g_{1/2}=122$,
$g_1=24$, $g_{3/2}=2$, $g_2=2$.}. %~\cite{NB1}
Toward the end of the evaporation
process, the temperature increases without apparent bound, although
the standard analysis breaks down at $T\sim\mpl$ (see Ref.~\cite{DV99}
for a discussion of the end point of evaporation).  Once the black
hole temperature $T\gg \mp$, moduli can be considered as
massless. Then the number of moduli emitted per PBH, with momentum $k$
between $k$ and $k+dk$, and per unit of time, is, for a Schwarzschild
black hole~\cite{H74}: $q_\phi(k,t) = (2\pi)^{-1}
\Gamma_\phi(M(t),k)/\left[\exp(k/T(t)) - (-1)^{2s}\right]$. The
absorption coefficient $\Gamma_\phi$ is a non-trivial function of $M$,
$k$ and $s$ which has to be calculated numerically~\cite{P76}, and $s$
is the spin of $\phi$. As announced any PBH will thus produce moduli
at some point, and, moreover, these moduli will be produced over a
whole range in momentum. To give an example of the sensitivity of the
constraints on $\Yp$ on the time of decay: if $\phi$ decays into
photons, pair creation on the cosmic background (of temperature
$T_\gamma$) suppresses cascade photons whose energy $E\ga
m_e^2/22T_\gamma$; since $T_\gamma\simeq1\,{\rm MeV}\,(t/1\,{\rm
s})^{-1/2}$, at early times $\la10^4\,$s, the cut-off lies below the
threshold of deuterium photo-dissociation ($\sim2\,$MeV), and the
constraints on $\Yp$ are evaded, while at later times, the cut-off is
pushed above this threshold, and photo-dissociation becomes highly
effective.  Finally, since a modulus carrying momentum $k$ at cosmic
time $\tp$ will decay at time $t\sim\tp{\rm max}[(k/\mp)^{2/3},1]$, it
is necessary to follow $d\Yp/dk$ as a function of time. As an aside,
this will permit the calculation of $\Yp$ produced by PBHs such that
$\Tbh<\mp$.

This calculation is carried out below in the following limits. As a
first approximation, it is sufficient to assume that all $\phi$
particles are emitted at the same average energy, parametrized as
$\alpha T(t)$; $\alpha$ is a constant which depends on $s$, with
$\alpha\simeq2.8$ for $s=0$, $\alpha\sim 4$ for $s=1/2$, and
$\alpha\sim7-8$ for $s=3/2$~\footnote{The value of $\alpha$ for
$s=3/2$ is based on extrapolation of the results of Ref.~\cite{McGW90}
for other spins, while the fraction of luminosity emitted in spin
$s=3/2$ (noted $\xp$ in the following) is given in
Ref.~\cite{McGW90}. It does not seem that a detailed study of Hawking
radiation of gravitinos has ever been performed.  Here it is assumed
that the helicity states $\pm1/2$ and $\pm3/2$ of the gravitino are
produced with values of $\alpha$ and $\xp$ as quoted for generic spin
$s=1/2$ and $s=3/2$ respectively.}~\cite{McGW90}.%~\cite{NB2} 
This approximation suffices as the energy at peak flux corresponds to
the average energy to within $\simeq10$\%~\cite{McGW90}, and since the
injection spectrum cuts-off exponentially for $k> \alpha T$, and as a
power-law for $k< \alpha T$. The initial mass fraction of PBHs is
approximated to a delta function centered on $\Mbh$. Although recent
considerations tend to indicate otherwise~\cite{NJ98}, this remains a
standard and simple approximation; moreover, the extension of the
results to a more evolved mass fraction is easy to carry out. Finally,
it is also assumed that the Universe is radiation dominated all
throughout the evaporation process, which implicitly implies that
black holes never dominate the energy density. This latter assumption
will be justified in Section 3.

Then the distribution $f_\phi(k,t)\equiv s^{-1} dn_\phi/dk = d\Yp/dk$,
where $s$ denotes the radiation entropy density, at times $\tbh<t<\tp$
reads:

\beq
f_\phi(k,t)= Y_{\rm BH} \int_{t_i}^{\tbh}q_\phi(k',t')\frac{dk'}{dk}dt'.
\label{Eq_fk}
\eeq

In this expression, $q_\phi(k',t')$ is the injection spectrum per
black hole as above, $Y_{\rm BH}\equiv n_{\rm BH}/s$, where $n_{\rm
BH}$ represents the PBH number density, and $k'\equiv k a(t)/a(t')$,
where $a$ is the scale factor. The factor $dk'/dk$ results from
redshifting of $k'$ at injection time $t'$ down to $k$ at time
$t$. Equation~(\ref{Eq_fk}) can be derived as the solution of the
transport equation: $\partial_t f_\phi = H\partial_k (kf_\phi) +
Y_{\rm BH}q(k,t)$, where $H$ is the Hubble scale at time $t$, and the
first term on the r.h.s accounts for redshift losses. This equation
and its solution Eq.~(\ref{Eq_fk}) are valid for $t\ll\tp$, when the
decay of $\phi$ particles can be neglected. It should be recalled that
in the range of masses $\mp$ and $\Mbh$ considered, indeed $\tbh \ll
\tp$. Equation~(\ref{Eq_fk}) also neglects the entropy injected in the
plasma by PBH evaporation, which remains a good approximation as long
as PBHs carry only a small fraction of the total energy density at all
times.

For mono-energetic injection at $k'=\alpha T(t')$:

\begin{equation}
q_\phi(k',t')= \frac{\xp}{\alpha T(t')}
\left|\frac{dM}{dt'}\right| \delta[k' - \alpha T(t')].\nonumber
\end{equation}

Here $\xp$ denotes the fraction of PBH luminosity
$\left|dM/dt'\right|$ carried away by moduli; for the MSSM content,
$\xp\simeq6\times10^{-3}$ for $s=0$ with one degree of freedom (e.g.,
a modulus field), $\xp\simeq6\times10^{-3}$ for $s=1/2$ with two
degrees of freedom (e.g., helicity $\pm1/2$ states of the gravitino),
and $\xp\simeq9\times10^{-4}$ for $s=3/2$ with 2 degrees of freedom
(e.g., helicity $\pm3/2$ states of the gravitino)~\cite{McGW90} (see
also previous footnote). %~\cite{NB2}
The $\delta$ distribution can be rewritten as
a function of $t$, using the identity: $\delta[f(t)]=|df/dt|^{-1}
\delta(t-t_s)$, where $t_s$ is such that $f(t_s)=0$ (here $t_s$ is
uniquely and implicitly defined in terms of $k$,
$k'$). Equation~(\ref{Eq_fk}) can be integrated in the limits $k\ll
k_0$ and $k\gg k_0$, where $k_0=\alpha\Tbh (t/\tbh)^{-1/2}$ is the
momentum at time $t$ of a particle injected at time $\tbh$ with
momentum $\alpha\Tbh$. In particular, modes with $k \ll k_0$ were
injected with energy $\simeq\alpha \Tbh$ at time $t'\simeq t
(k/\alpha\Tbh)^2 \ll \tbh$, while modes with $k \gg k_0$ were produced
in the final stages at time $t'\simeq\tbh$ with momentum $k'\simeq
\alpha T(t')\gg \alpha\Tbh$. One obtains:

\begin{eqnarray}
f_\phi(k,t) & \simeq & \frac{2}{3}\xp\frac{\Mbh}{(\alpha\Tbh)^2}
\frac{k}{\alpha\Tbh}\frac{t}{\tbh} Y_{\rm BH} \,\,\, (k\ll k_0),
\label{Eq_fkm}\\
f_\phi(k,t) & \simeq &
\xp\frac{\Mbh}{(\alpha\Tbh)^2}\left(\frac{k}{\alpha\Tbh}\right)^{-3}
\left(\frac{t}{\tbh}\right)^{-1}Y_{\rm BH}\,\,\, (k\gg k_0),
\label{Eq_fkp}
\end{eqnarray}
and both expressions agree to within a factor $3/2$ at $k=k_0$. 

If initially $\Tbh < \mp$, moduli are produced only in the final
stages for $k' \gg \mp$ at injection. Hence the above spectrum should
remain valid if a low--momentum cut-off $k_c\sim \mp (t/\tbh)^{-1/2}>
k_0$ is introduced. The total number of $\phi$ particles produced
(hence the constraint on $\beta$) is thus suppressed (weakened) by a
factor $\sim (\mp/\Tbh)^2$, after integration of $f_\phi(k,t)$ over
$k>k_c$, if $\Tbh<\mp$, i.e., if $\Mbh> 10^9\,{\rm g}\,(\mp/10\,{\rm
TeV})^{-1}$. Since this mass range $\Mbh\ga10^9\,$g is moreover
strongly constrained by the effects on BBN of quarks directly produced
in the evaporation~\cite{KY99}, it will be ignored in the following.

For PBHs such that $\Tbh\ga\mp$, it is a very good approximation to
consider that emitted moduli carry at time $t$ a momentum $k_0$, since
$kf_\phi(k,t)=d\Yp/d\ln(k)$ behaves as $k^2$ for $k\ll k_0$, and as
$k^{-2}$ for $k\gg k_0$. Moreover at time $t=\tp$:

\beq \frac{k_0}{\mp} \simeq 0.01\alpha\left(\frac{\mp}{1\,{\rm
TeV}}\right)^{1/2} \left(\frac{\Mbh}{10^9\,{\rm g}}\right)^{1/2}, \eeq

and therefore the $\phi$ particles decay at rest (in the plasma rest
frame), at time $\tp\sim \Mpl^2/\mp^3$, in the range of masses
considered, $\mp\la 10\,$TeV and $\Mbh\la10^9\,$g. One then seeks the
total number of moduli present at that time, which is given by:

\beq \Yp\simeq \frac{\xp}{2}\frac{\Mbh}{\alpha\Tbh} Y_{\rm BH}.
\label{Eq_Yp}
\eeq

This result can be obtained as a solution of the transport
equation\linebreak $\partial_t \Yp = \xp Y_{\rm BH}
\left|dM(t)/dt\right|/\alpha T(t)$, or by integrating $f_\phi(k,t)$
over $k$ in Eqs.~(\ref{Eq_fkm}), (\ref{Eq_fkp}) above; all three
results agree to within a factor $3/2$. Equation~(\ref{Eq_Yp}) has a
simple interpretation: within a factor $2$ it corresponds to the
instantaneous evaporation of black holes at time $\tbh$, with total
conversion of their mass $\Mbh$ in particles of energy $\alpha\Tbh$, a
fraction $\xp$ of which is moduli. This result can be rewritten in
terms of more conventional parameters. The mass $\Mbh$ is taken to be
a fraction $\delta$ of the mass within the horizon at the time of
formation $t_i$: $\Mbh \approx 4\pi\delta \mpl^2/H_i$, where $H_i$
denotes the Hubble scale at time $t_i$, and $\delta\sim{\cal O}(1)$ is
expected~\cite{NJ98}. Furthermore, instead of $Y_{\rm BH}$, one
generally uses the mass fraction $\beta\equiv n_{\rm BH}\Mbh/\rho_{\rm
c}$ defined at the time of PBH formation $t_i$, with $\rho_{\rm
c}=3H_i^2\mpl^2$ the critical energy density at that time. Using
$s=(2\pi^2/45)g_\star T_\gamma^3$, with $T_\gamma$ the cosmic
background temperature, $T_\gamma\approx 0.5 g_{200}^{-1/4}
H_i^{1/2}\mpl^{1/2}$, and $g_{200}=g_\star/200$ ($g_\star$ number of
degrees of freedom), one obtains:

\beq 
\beta\simeq 3\times10^{21} g_{200}^{1/4}\delta^{-1/2}
\left(\frac{\Mbh}{10^9\,{\rm g}}\right)^{3/2}Y_{\rm BH}, 
\eeq 
and therefore:

\beq
\Yp\simeq 2\times10^4\delta^{1/2}g_{200}^{-1/4}
\left(\frac{\xp}{6\,10^{-3}}\right)\left(\frac{\alpha}{3}\right)^{-1}
\left(\frac{\Mbh}{10^9\,{\rm g}}\right)^{1/2}\beta,
\label{Eq_Ypf}
\eeq 
which constitutes the main result of this section. If the Universe
went through a matter dominated era between times $t_a$ and $t_b$,
with $t_i<t_a<t_b\ll\tbh$, then the r.h.s. of Eq.~(\ref{Eq_Ypf}) must
be multiplied by the factor $(H_b/H_a)^{1/2}$, where $H_{a,b}$ is the
Hubble scale at time $t_{a,b}$, and the constraint on $\beta$
Eq.~(\ref{Eq_beta}) below is weakened consequently.

\bigskip

%\vspace{1cm}

%\noindent{\bf 3. Discussion}
%\vspace{0.3cm}
%\section{Discussion}

{\it 3. Discussion --} As mentioned previously, the most stringent
constraints on $\Yp$ result from the effect of the decay products of
$\phi$ on BBN~\cite{ELN82}. These studies assume monoenergetic
injection at energy $\mp$ at time $\tp$, and their results can be used
safely, since the moduli emitted by PBHs decay when non-relativistic.
One usually considers production of hadrons or photons in $\phi$
decay. The constraint due to hadron injection is in principle very
significant for $\mp\ga1\,$TeV, but it is not obvious that $\phi$ can
decay hadronically, and moreover it relies on assumptions on the
cosmic evolution of helium-3 (see, e.g., Ref.~\cite{ELN82}), which
have now been proven uncertain (see, e.g., Ref.~\cite{OSW99} and
references therein). Therefore, in the following, only constraints on
photon injection are used; the bounds presented will thus be slightly
conservative. Holtmann {\it et al.}~\cite{ELN82} obtain in this
case:\medskip

$\Yp\la10^{-15}$ for $\mp\simeq100\,\GeV$, 

$\Yp\la10^{-14}$ for $\mp\simeq300\,\GeV$,

$\Yp\la5\times10^{-13}$ for $\mp\simeq1\,$TeV, and 

$\Yp\la5\times10^{-10}$ for $\mp\simeq3\,$TeV. \medskip

The error on the upper limit is a factor $\simeq4$. It results from
the uncertainty in the fudge factors that enter the $\tp(\mp)$
relationship when the constraints of Holtmann {\it et al.}, given in
the plane $\mp\Yp$--$\tp$, are translated into the plane
$\Yp$--$\mp$. Note that these constraints assume that $\phi$ decays
into photons with a branching ratio unity, and should be scaled
consequently. However these limits should also be strengthened by
a factor as high as $\sim50$ to avoid $^6$Li overproduction, if one
assumes that $^6$Li has not been destroyed in stars in which it has
been observed~\cite{J99}. This constraint is ignored in what follows,
as it relies on yet unproven assumptions on stellar evolution; this
but makes the above constraints more conservative. Finally, the
observational upper limit on the amount of $\mu-$distortion in the
cosmic microwave background implies~\cite{NOS83,ELN82}: $\Yp \la
10^{-13}(\mp/100\,{\rm GeV})^{1/2}$ for $20\,\GeV\la\mp\la500\,\GeV$.

Note that the most stringent constraint on $\beta$ results from the
production of the lightest of all moduli--like particles in the
theory, whose mass would likely be $\la {\rm
few}\times100\,$GeV. Overall it seems that $\Yp\la10^{-14}$ represents
a reasonable generic upper limit from BBN. Using Eq.~(\ref{Eq_Ypf}),
this can be rewritten as a limit on $\beta$:
%\begin{eqnarray}
\begin{equation}
\beta 
%& \la & 
\la 
5\times10^{-19} \delta^{-1/2}g_{200}^{1/4} 
 \left(\frac{\xp}{6\,10^{-3}}\right)^{-1}
 \left(\frac{\alpha}{3}\right) 
%\nonumber\\ & & \times 
\left(\frac{\Mbh}{10^9\,{\rm
 g}}\right)^{-1/2} \left(\frac{\overline\Yp}{10^{-14}}\right),
\label{Eq_beta}
\end{equation}
%\end{eqnarray}
and $\overline\Yp$ denotes the upper limit on $\Yp$.  This result does
not depend on whether PBHs are shrouded in a photosphere, as suggested
by Heckler~\cite{H97}, since moduli are not expected to interact with
it due to their gravitational interaction cross-section. On the
contrary, other astrophysical constraints on $\beta$ for
$\Mbh\ga10^9\,$g are in principle sensitive to the presence of a
photosphere, as they rely on the direct emission of photons and
quarks~\cite{CMcG98}.

If $R-$parity holds, the constraint on the LSP mass density
$\Omega_{\rm LSP}<1$ today implies:\linebreak $\beta \la
2\times10^{-17}\delta^{-1/2}g_{200}^{1/4} (\alpha/3) (x_{\rm
LSP}/0.6)^{-1} (\Mbh/10^9\,{\rm GeV})^{-1/2} (m_{\rm LSP}/100\,{\rm
GeV})^{-1}$. This constraint has been adapted from the study of
Ref.~\cite{G99} and Eq.~(\ref{Eq_Ypf}) above. The fraction of
luminosity carried away by the LSP is $x_{\rm LSP}\simeq 0.6$, since
each spartner produced by a PBH will produce at least one LSP in its
decay~\cite{G99}. This LSP constraint on $\beta$ is thus less
stringent than the moduli constraint, provided at least one modulus of
the theory has mass $\la 1\,$TeV.

These results have several implications. First of all, the
approximation made in Sec.~2, namely $\Omega_{\rm BH}\ll1$ at all
times is justified. In effect, $\Omega_{\rm BH}=\beta (t/t_i)^{1/2}$
at time $t$ in a radiation-dominated Universe, since PBHs behave as
non-relativistic matter, and therefore at time $\tbh$, $\Omega_{\rm
BH}\simeq 2.3\times10^{14}\delta^{1/2}(\Mbh/10^9\,{\rm
g})\beta$. Consequently, if $\beta$ verifies the above upper limits,
indeed $\Omega_{\rm BH}\ll1$ at all times. However, since
Eq.~(\ref{Eq_Ypf}) is not valid if $\Omega_{\rm BH}=1$ at some time
$t_\star<\tbh$, one needs to consider this case as well.

An order of magnitude of $\Yp$ in this case can be obtained as
follows. If $t_\star\ll\tbh$, the radiation present subsequent to PBH
evaporation has been produced in the evaporation process itself.
Assuming total conversion of the PBH mass $\Mbh$ at time $\tbh$ into
particles (moduli and radiation) of energy $\alpha\Tbh$, one finds
$n_\phi\approx \xp \rho_{\rm BH}/\alpha \Tbh$, and $s\approx
(4/3)\rho_{\rm BH}/T_{\rm RH}$, where $\rho_{\rm BH}$ is the PBH
energy density at evaporation, $T_{\rm RH}\approx 2\,{\rm
MeV}\,g_{10}^{-1/4} (\Mbh/10^9\,{\rm g})^{-3/2}$ is the reheating
temperature, and $g_{10}=g_\star/10$. Therefore $\Yp \approx 3\times
10^{-10} g_{10}^{-1/4}(\xp/6\times10^{-3})(\alpha/3)^{-1}
(\Mbh/10^9\,{\rm g})^{-5/2}$, well above the previous limits. Note
that one would naively expect $\Yp\sim\xp$ since $\xp$ is the fraction
of PBH luminosity carried away by $\phi$ particles. However the
photons emitted by PBHs carry high energy $\simeq \alpha \Tbh$ and small
number density $\sim \rho_{\rm BH}/\alpha\Tbh$, and their
thermalization leads to many soft photons carrying high
entropy. Nevertheless, this discussion shows that PBHs of any mass
should never come to dominate the energy density; if this were to
happen, PBHs with $\Mbh\la10^9\,$g would produce too many moduli,
while the evaporation of PBHs with $\Mbh\ga10^9\,$g would lead to too
low a reheating temperature. In particular, scenarios of reheating of
the post-inflationary Universe by black hole evaporation, as put
forward, e.g., in Ref.~\cite{GBLW96}, are forbidden. This result
was also envisaged in Ref.~\cite{FKL99}.

Finally, the present constraints on $\beta$ exclude the possibility of
generating the baryon asymmetry of the Universe through PBH
evaporation. Indeed Barrow {\it et al.}~\cite{C76} have performed a
detailed computation of the baryon number to entropy density ratio
$n_{\rm b}/s$ produced in PBH evaporation, and find: $n_{\rm
b}/s\simeq 7\times10^4 \epsilon (x_{\rm H}/0.01) g_{200}^{-1/4}
(\Mbh/10^9\,{\rm g})^{1/2}\delta^{1/2}\beta$, where $\epsilon$ is the
baryon violation parameter, defined as the net baryon number created
in each baryon-violating boson decay, $x_{\rm H}$ is the fraction of
PBH luminosity carried away by such bosons, and other notations are as
above. Unless all moduli--like particles are heavier than
$\sim3\,$TeV, and $R-$parity does not hold, the above constraints on
$\beta$ imply $n_{\rm b}/s< 10^{-12}\epsilon$, which does not suffice
since BBN indicates $n_{\rm b}/s\sim4-7\times10^{-11}$.

\end{document}